\documentclass{easychair}

\usepackage{doc}
\graphicspath{ {./ProjectImages/} }

\usepackage{titlesec}
\usepackage{array}
\usepackage{enumerate}
\usepackage{multirow}
\usepackage{amsmath}
\usepackage{subfig}
\usepackage{pgfplots}
\usepackage{scalefnt}
\usepackage{tikz}

\title{Developing and experimenting with LEO satellite constellations in OMNeT++}

\author{
Aiden Valentine
\and
    George Parisis
}

\institute{
  School of Engineering and Informatics\\
  University of Sussex, UK\\
  \email{av288@sussex.ac.uk}, \email{G.Parisis@sussex.ac.uk}
 }

\authorrunning{A.Valentine and G.Parisis}

\titlerunning{Developing and experimenting with LEO satellite constellations in OMNeT++}

\begin{document} 

\maketitle

\begin{abstract}
    In this paper, we present our work in designing and implementing a LEO satellite constellation simulation model\footnote{The LEO satellite constellation simulation model can be found within the following GitHub repository:  https://github.com/Avian688/leosatellites} within OMNeT++ and INET, which is validated by comparing the results with existing work. Our model builds upon the fundamentals of the Open Source Satellite Simulator (OS$^3$ ), which was ported to INET 4.3. We describe how the model was integrated on top of the INET and ported OS$^3$  Framework. We then experiment with the simulation model to demonstrate its viability in simulating LEO satellite constellations. This involved simulating both outdated and more recent satellite constellations, using FCC filing information, to validate latency results.
\end{abstract}

%
%

\section{Introduction}
\label{sect:introduction}

	A new space race established due to the thirst for higher internet speeds and global coverage is currently underway.  SpaceX \cite{spacexStarlink} is competing against a wide variety of private companies \cite{LEOSatellitesTelesat, OneWebOneWorld} to deploy Low Earth Orbit (LEO) satellite constellations \cite{klenzeNetworkingHeavenEarth2018}, promising a revolution in internet technology. It is vital that sufficient research is undertaken as the deployment of LEO satellite constellations will essentially set up the future backbone of the Internet. These LEO satellite constellations will exhibit distinctive characteristics that set it apart from today’s fibre. This includes an interconnected dynamic mesh of thousands of satellites \cite{handleyDelayNotOption2018}, substantial latency reductions in comparison to the best fibre available today \cite{bhattacherjeeGearing21stCentury2018},  an aggregate capacity expected to reach multiple Tbps and an average round-trip-time in a sub-10ms range between the first satellite and the Earth \cite{klenzeNetworkingHeavenEarth2018}.
	To allow thorough experimentation of networks, simulation models are fundamental. Existing research has been predominately based on the use of custom simulation models, which are not available to the public, making the validity of these results difficult to verify. Open-Source simulation platforms, such as OMNeT++ \cite{OMNeT}, are publicly available, allowing easy replication of results. OMNeT++ is an object-orientated modular network simulation framework that provides the foundation and tools for writing network-based simulations. Alongside the INET framework \cite{INETFrameworkINET}, many models are openly available which can accurately simulate the Internet stack. Its simulation models are constructed by modules - which are components allowing message passing through gates and connections. OMNeT++ is an ideal simulation platform for the development of satellite networks, which has been shown by the development of the Open Source Satellite Simulator (OS$^3$) \cite{niehoeferCNIOpenSource2013}. As large LEO satellite constellations are yet to be accomplished, simulation models are even more important; allowing researchers to test the effectiveness of protocols in the satellite constellations before they are deployed. This enables potential rethinks of protocols and mechanisms, possibly accelerating the long-term goal of global high-speed internet. We present a simulation model based on OMNeT++/INET for evaluating the latencies for LEO satellite constellations.

\section{Starlink}
\label{sect:starlink}

There are currently 1,630 LEO satellites orbiting the earth which are all part of SpaceX’s vast plan for global high-speed internet access. In 2018 SpaceX had their proposal for Starlink, the global LEO satellite constellation which is already making monumental progress, accepted by the US Federal Communications Commission (FCC) \cite{SpaceXNongeostationarySatellite}. With nearly 1,700 satellites already deployed by SpaceX, global internet coverage is highly likely to arrive within the decade. 

SpaceX initially set out with the intention of deploying 4425 satellites with different deployment stages, each stage promising high-speed internet access to specific areas.  Since this initial intention, the US FCC has approved further 42,000 Starlink satellites, which will operate at lower altitudes from 328 to 580 km in comparison to the previous 1,100 to 1,325 km range \cite{FCCFormAttachment}. As noted by the FCC application submitted by SpaceX \cite{FCCFormAttachment}, this altitude change reduce the potential propagation delays between ground stations and satellites, as well as reduce the debris collision risk that the constellations presented \cite{boleySatelliteMegaconstellationsCreate2021}. Collisions are not a significant consideration of the simulation model, as even though they are vital, the model focuses on the LEO constellations strictly from a network perspective. 

Due to the public knowledge provided by SpaceX and research undertaken in comparison to other private companies (such as Telesat \cite{LEOSatellitesTelesat} and OneWeb \cite{OneWebOneWorld}), we used Starlink as the primary resource for the specification behind the LEO satellite constellation simulation model. FCC filings for Starlink \cite{SpaceXNongeostationarySatellite} provided sufficient information to make a realistic model.
\section{Open Source Satellite Simulator -  OS$^3$ }
\label{sec:os3}

To simulate a dynamic constellation of satellites the mobility of a node is vital. Many aspects such as propagation delay, signal strength and interference are all impacted by the distance between nodes, making mobility the foremost step. The OS$^3$ framework provides suitable mobility modules that allow the importation of current satellite information into OMNeT++ and INET. These mobility models use the established SDP4 and SGP4 \cite{hootsSpaceTrackReportNo1980} mathematical algorithms to accurately simulate the positions of a satellite at a given time. We ported the outdated OS$^3$ framework for it to work appropriately with the current INET version 4 \footnote{The ported framework and changes can be found within the following GitHub repository: https://github.com/Avian688/os3}. These changes include, adapting the ground station mobility model \textit{LUTMotionMobility} from extending the INET 2.5 \textit{MobilityBase} model to the more current \textit{StationaryMobility} model and adapting the satellite mobility model \textit{SatSGP4Mobility} to extend the current INET \textit{MovingMobilityBase} rather than \textit{MobilityBase} model. The primary porting efforts were focused on porting only the mobility models of OS$^3$ into the latest OMNeT++ and INET. OS$^3$ also includes a graphical user interface for easy simulation, but this does not allow the integration of new models and was consequently not the focus.

\section{LEO Satellite Constellation Simulation Model }
\label{sec:simulationModel}
Our ground-station and satellite models both extend the INET \textit{WirelessHost} model. This model defines a host with wireless capabilities which is compatible with INETs vast array of models and protocols, providing sufficient components that allow an accurate simulation of the TCP/IP stack in the process. Ground-to-satellite phased array beams and inter-satellite links are extremely complex, therefore many factors can be considered when transmitting and receiving signals within the simulation model. Like existing work \cite{handleyDelayNotOption2018}, the initial simulation model and experiments predominately focus on latency which is only limited by the topology and speed of light. As a consequence of being built within the OMNeT++/INET framework, increasing the fidelity of the simulation model is very straightforward as additional modules can be integrated without inconvenience. By using the \textit{WirelessHost} model, the additionally developed models defined below are simply integrated, through either the NED language or specified within the network INI file.
\paragraph{Physical Layer}
To keep up with the dynamic mesh of satellites, we decided that the INET \textit{UnitDiskRadio} and \textit{UnitDiskRadioMedium} models were suitable \footnote{A more complex model such as INET ApskScalarRadio was considered, but was ultimately not used due to the desire to re-evaluate existing work \cite{handleyDelayNotOption2018, handleyUsingGroundRelays2019, bhattacherjeeGearing21stCentury2018, klenzeNetworkingHeavenEarth2018}, which focuses on latency only limited by topology and the speed of light.}. These INET physical layer models provide fast and predictable physical layer behaviour, which is a perfect candidate for a large scale moving network. The \textit{UnitDiskRadio} model is based on three parameters, a communication range, interference range and a detection range. The interference of the model is ignored as this is not relevant to the fidelity of the initial simulation model. We implemented new INET physical layer models that use the longitude, latitude and altitude of the OS$^3$ mobility models. This includes new UnitDisk models (\textit{SatelliteUnitDiskTransmitter} and \textit{SatelliteUnitDiskTransmission}) and a propagation model (\textit{SatellitePropagation}). The OS$^3$ mobility modules map only the longitude and latitude (altitude is ignored) onto the OMNeT++ X, Y and Z coordinates, most likely for visualisation purposes. There are many issues with projecting the Earth onto a two-dimensional plane \cite{delaneyProblemProjections} therefore by creating new INET models that only use the longitude, latitude and altitude, the calculation of distance for determining propagation delay between nodes is much more accurate.
\paragraph{Routing}
We implemented a new INET configurator module, \textit{SatelliteNetworkConfigurator}, which follows the \textit{IPv4NetworkConfigurator} model in INET. The \textit{IPv4NetworkConfigurator} works in INET as follows: 1) A graph is built representing the current network topology, 2) IP addresses are assigned, 3) Link and node weights are set, 4) Dijkstra's Shortest Path Algorithm is run from each network node, where all the routes are determined, 5) The routing tables for each node are generated. The \textit{SatelliteNetworkConfigurator} has been created with the intent to run throughout the simulation, thus a new method (\textit{SatelliteNetworkConfigurator::reinvokeConfigurator}) ensures that the allocated memory for the past topology is released, and then goes through the configuration steps again to rebuild a new topology. Step two is ignored once the configuration is reinvoked, as the IP addresses have already been assigned, and it would be redundant to reassign them. The frequency that the configurator recreates a new topology is determined by the INI file parameter \textit{updateInterval}. The \textit{SatelliteNetworkConfigurator} also filters links within step 3, to ensure that links are valid. This includes whether or not satellites are compatible for inter-satellite links or if a ground-satellite link satisfies the minimum elevation angle stated within the Starlink FCC filings \cite{FCCFormAttachment}.

\begin{figure}[!htb]
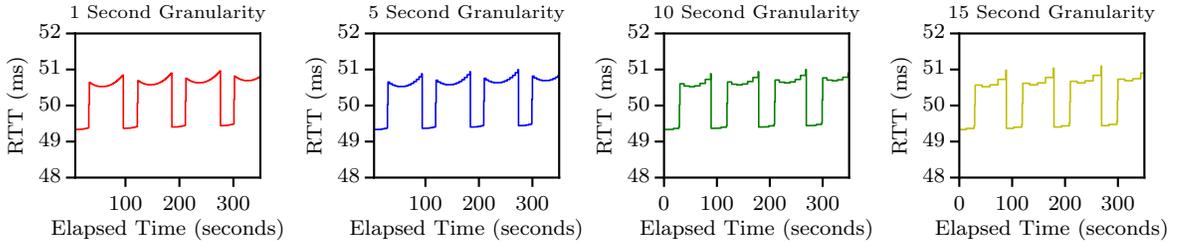
%
	\centering
	\subfloat[\centering 1 Second Granularity - (Elapsed Real Time: 15m 55s)]{{\input{./ProjectImages/DC-FrankfurtGranularityOneSec.pgf} }}%
	\subfloat[\centering 5 Second Granularity - (Elapsed Real Time: 6m 26s)]{{\input{./ProjectImages/DC-FrankfurtGranularityFiveSec.pgf}}}%
	\subfloat[\centering 10 Second Granularity - (Elapsed Real Time: 4m 18s)]{{\input{./ProjectImages/DC-FrankfurtGranularityTenSec.pgf}}}%
	\subfloat[\centering 15 Second Granularity (Elapsed Real Time: 3m 55s)]{{\input{./ProjectImages/DC-FrankfurtGranularityFifteenSec.pgf}}}%
	\caption{One to Fifteen Second Routing Granularity. 660 Inter-Satellite Link Satellites are being simulated.}%
	\label{fig:1-15GranularityResults}%
\end{figure}

Figure \ref{fig:1-15GranularityResults} shows how altering the \textit{updateInterval} parameter impacts the latency between two ground-stations using the simulation model. All of the experiments are tested on a constellation of 660 satellites with inter-satellite links, with pings being sent between Washington DC to Frankfurt. The constellation mirrors the 2018 FCC specification \cite{FCCForm2018}, where there is 66 satellite per plane with an inclination of $53^{\circ}$. Figures \ref{fig:1-15GranularityResults}c and  \ref{fig:1-15GranularityResults}d both indicate a slight degradation of results when there is a lower granularity of the \textit{updateInterval} parameter. Despite the minor loss in accuracy, the results still show that a lower granularity is competent in showing the patterns of RTT changes due to the paths moving to an alternate orbital plane. The runtime of Figure \ref{fig:1-15GranularityResults}a shows that the one-second update interval is noticeably slower in comparison to the other experiments. With larger constellation sizes of 1500 satellites, the runtime will be even larger, with the configurator taking up to a minute to reconfigure the routes.

\begin{figure}[!htb]
	\centering
	\input{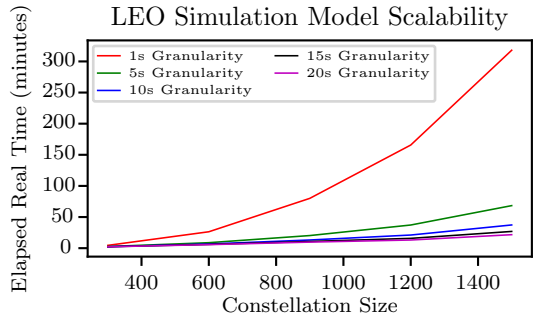}
	\caption{Scalability with varying constellation sizes. All of the experiments are tested on a constellation with 24 orbital planes, with 66 satellites per plane. The simulation is simulating five minutes of the constellation sending ping requests every 500ms from Los Angeles to New York. Inter-satellite links are enabled for the constellation.}
	\label{fig:Scalability}
\end{figure}

Figure \ref{fig:Scalability} is a graph showing the scalability of the simulation model. As the size of the satellite constellation increases, the \textit{SatelliteNetworkConfigurator} will need to build an even larger graph representing the network topology. This results in a higher routing complexity and larger routing tables. Extremely large constellation sizes can take the simulation model up to five hours of real-time to simulate five minutes of simulation time with the \textit{SatelliteNetworkConfigurator} routing every 1 second using a 3.2 GHz Intel Core i7 processor. The run time of the simulation drastically decreases when increasing the update interval of the routing, at the cost of varying levels of accuracy loss (see Figure \ref{fig:1-15GranularityResults}) depending on the granularity.

\paragraph{Removing the TLE data set dependency}
A large limitation of the simulation model is that the OS$^3$ satellite mobility models require existing satellite data (Two-Line Element sets). However, this presents many problems such as the inability to simulate unimplemented/experimental satellite constellations and the need to constantly update the TLE data files as more satellites are deployed. Another issue is that the satellite data provides no information on inter-satellite connectivity, making implementing the laser links within the simulation model difficult. We decided that to be able to replicate existing inter-satellite latency experiments as well as to enable the capability of exploring any future constellation, a new \textit{Norad} module was implemented that does not depend on two-lined element set data.

This module instead automatically generates specific orbital elements depending on a combination of Keplerian elements and constellation characteristics that are specified within the network INI file. A phase offset is also considered \cite{handleyDelayNotOption2018}, which is the offset between satellites on adjacent orbital planes to prevent collisions. This value is included within the NoradA module to allow more precise satellite constellation deployments if required.

For ease of use within the developed simulation model, the original OS$^3$ \textit{Norad} module was split into three different modules. This includes the module interface \textit{INorad}, and two sub-modules \textit{NoradA} and \textit{NoradTLE}. \textit{NoradTLE} is identical in implementation to the original OS$^3$ \textit{Norad} module, where existing satellite data is used to propagate the orbits of the satellites. The \textit{NoradA} module demonstrates the new automatic generation implementation, where different orbital parameters can be specified by the NED language or INI file. 

\begin{figure}[!htb]%
	\centering
	\subfloat[\centering The constellation has been defined in the INI file with an inclination of $53^{\circ}$, 66 satellites per plane, and a single plane.]{{\includegraphics[width=6cm]{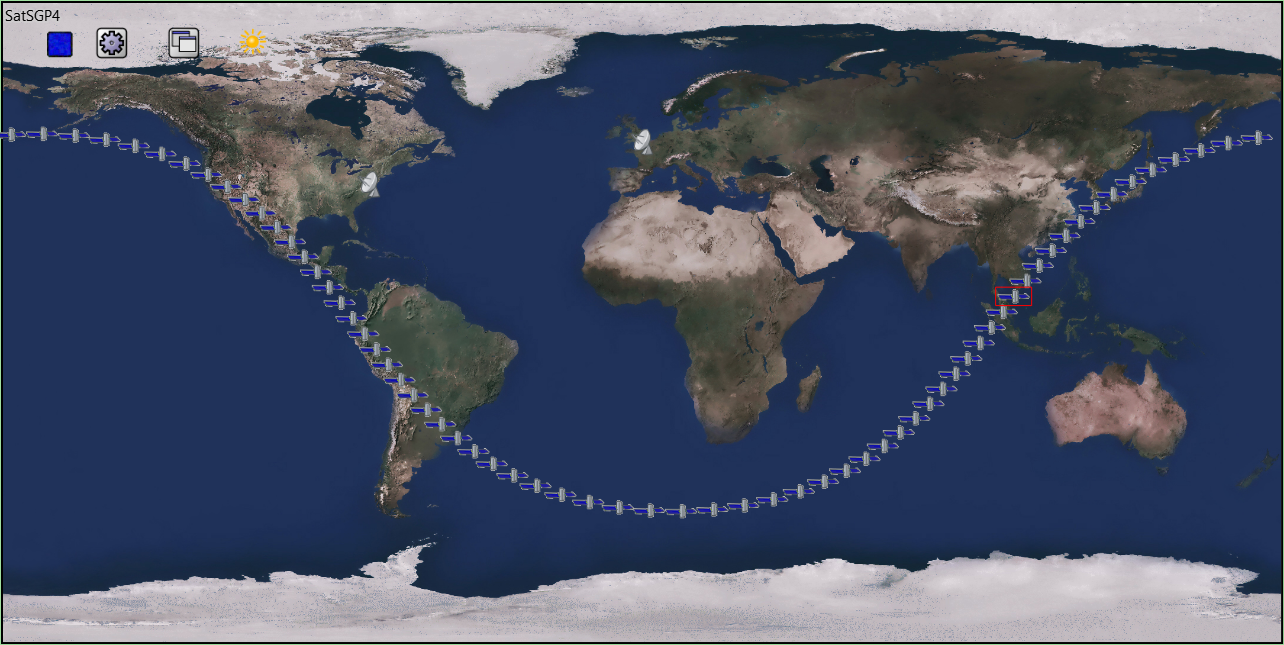} }}%
	\qquad
	\subfloat[\centering The entire first phase is simulated, with satellites having a $53^{\circ}$ inclination, 72 satellites per plane, and 22 planes.]{{\includegraphics[width=6.5cm]{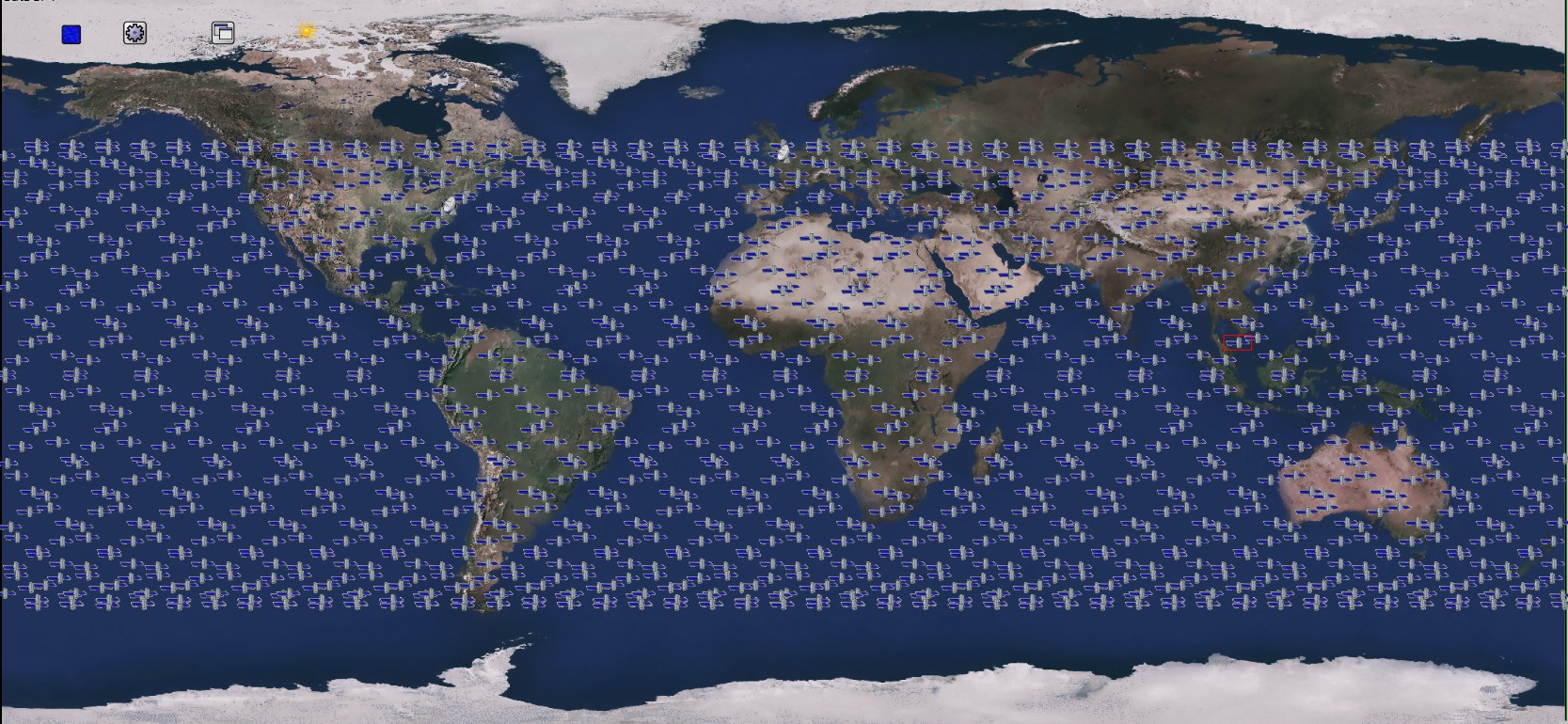} }}%
	\caption{Satellite constellations using the custom \textit{NoradA} module with parameters defined by the latest Starlink FCC filings \cite{FCCFormAttachment}. }%
	\label{fig:constellationNorad}%
\end{figure}

\paragraph{Inter-Satellite Links}
As each satellite is given an index position from 0 to the maximum satellite number specified, which is used to determine its orbital plane and position, it is possible to determine inter-satellite links. Starlinks FCC filings \cite{FCCFormAttachment} state that each satellite will have four inter-satellite links. \cite{handleyDelayNotOption2018} notes that for the best connectivity, satellites will most likely connect to the two satellites on either side of its current orbital plane, and the two adjacent satellites that are on neighbouring orbital planes. As each satellite has an index within the simulation model, it is extremely easy to determine the four laser links of a satellite. A method was created within the \textit{NoradA} class (\textit{NoradA::isInterSatelliteLink})which determines whether, given a satellites index, it is compatible as an inter-satellite link. This method is used during the link filtering within the \textit{SatelliteNetworkConfigurator} so that any non-compatible inter-satellite links are disabled.
\section{Experiments}
\label{sec:experiments}
To validate the simulation model we conducted a variety of experiments, mirroring existing work \cite{handleyDelayNotOption2018, handleyUsingGroundRelays2019, bhattacherjeeGearing21stCentury2018}. INETs \textit{PingApp} module will be used to simulate ping between two ground stations so that the round-trip time between two locations can be examined and compared to existing work results. The \textit{PingApp} will generate ping requests from an initial ground station to another ground station every 500ms throughout the simulations.
\paragraph{Experiment 1}
The first simulation that we evaluated mirrored the specification defined in \cite{handleyUsingGroundRelays2019}. The constellation involves 66 satellites per plane, with a total of 24 planes, which are generated using the \textit{NoradA} module. There are 14 ground relays used in this simulation, with two being situated on ships between the source and destination; London and New York. The inter-satellite link experiment will update the mobility and routing every five seconds to reduce the run-time. This experiment will simulate 120 minutes of simulation time.
\begin{figure}[!htb]
	\centering
	\input{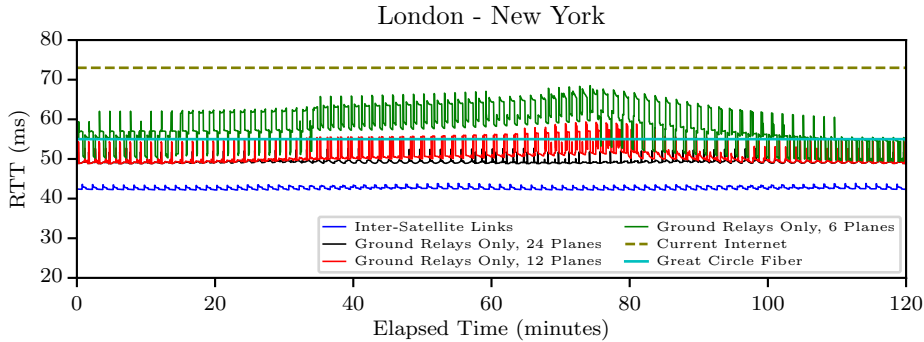}
	\caption{London - New York RTT comparing Inter-Satellite links and Ground Relays}
	\label{fig:London-NYResults}
\end{figure}

The results validate the simulation model by accurately recreating the results described in \cite{handleyUsingGroundRelays2019}. The ground relays that are only using 6 orbital planes provide varying latencies of 50-63 ms in comparison to the current Internet's 73 ms RTT \cite{GlobalPingStatistics}. The more planes, the lower the latencies with 12 and 24 planes rivalling an ideal Great Circle optical fibre \cite{CorningSMF28Ultra} \footnote{Fibre running in a straight line from source to destination}. The results also show that the inter-satellite links perform better than any of the comparisons, with a mean RTT of 43 ms. The variations of round trip times are a result of the shortest path between London and New York changing to a different orbital plane as a satellite becomes inaccessible, or a better path is emerged due to a new satellite becoming within range. As evident in the results more satellites lower the chances of a route becoming inaccessible due satellites going out of range, as well as providing potentially better paths that the lower constellations sizes were unable to provide. With ground relays, there are less potential routes depending on the amount of relay stations that are used, leading to more drastic fluctuations in RTT in comparison to inter-satellite links.

\begin{figure}[!htb]
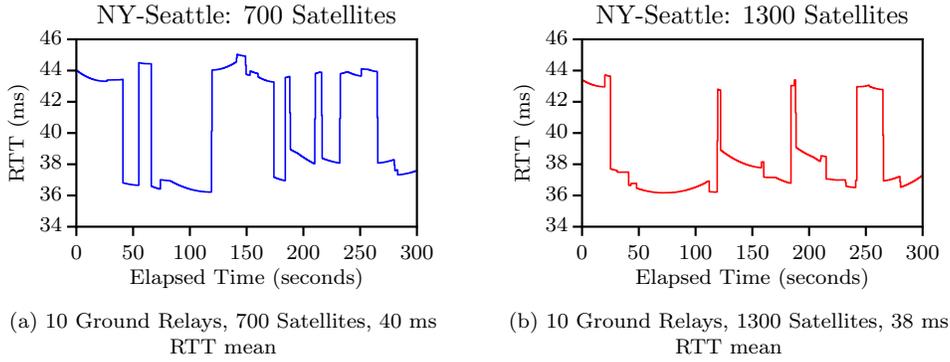
%
	\centering
	\subfloat[\centering 10 Ground Relays, 700 Satellites, 40 ms RTT mean]{\scalebox{1}{\input{./ProjectImages/TLEGroundRelay2.pgf}}}%
	\qquad
	\subfloat[\centering 10 Ground Relays, 1300 Satellites, 38 ms RTT mean]{\scalebox{1}{\input{./ProjectImages/TLEGroundRelay.pgf}}}%
	\caption{Current constellation latencies between New York and Seattle. TLE data sets are used.}%
	\label{fig:groundRelayResults}%
\end{figure}

\begin{figure}[!htb]
	\centering
	\input{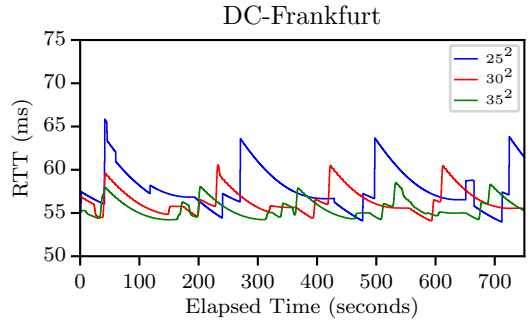}
	\caption{DC - Frankfurt RTT. Varying constellation sizes are used from $25^2$ to $35^2$.}
	\label{fig:DC-FrankfurtReevaluation}
\end{figure}

Figure \ref{fig:groundRelayResults} depicts another simulation example with information used from \cite{handleyDelayNotOption2018} and \cite{handleyUsingGroundRelays2019}. This simulation uses the TLE data sets of 1300 of the current Starlink constellation, demonstrating the \textit{NoradTLE} module in contrast to the previous simulation. The simulation portrays how two different sizes of satellite constellations affect the RTT between New York and Seattle using 10 ground relay stations that replicate the positions of existing Starlink control centres \cite{StarlinkSatelliteTrackera}. The results show that using the \textit{NoradTLE} module leads to less predictable fluctuations of RTT, which is due to SpaceX strategically placing its satellites rather than simply filling out an orbital plane at a time. This makes gaps within an orbital plane much more likely, causing more variations in RTT as paths will frequently shift between orbital planes. From 700 to 1300 satellites, the RTT begins to stabilise as there are more shortest paths available and less potential gaps within an orbital plane, which will continue for future Starlink deployments until the first phase is complete.
\paragraph{Experiment 2}
To test the capabilities of the simulation model, a completely uniform constellation with each orbital plane having an inclination of $90^{\circ}$ was simulated, shown in Figure \ref{fig:DC-FrankfurtReevaluation}. This simulation was juxtaposed with the results shown in \cite{bhattacherjeeGearing21stCentury2018}. The simulations consist of constellation sizes of $N^2$ where there are N orbital planes with N satellites per plane. Three constellation sizes will be evaluated where $N={25,30,35}$. A constellation size of $20^2$ was not tested unlike \cite{bhattacherjeeGearing21stCentury2018} as the minimum elevation angle specified by FCC filings \cite{FCCFormAttachment} makes them infeasible for solely inter-satellite links. 

The simulation also differs by using a mean altitude of 550 km for satellites to reflect recent FCC filings \cite{FCCFormAttachment}. The results demonstrate the advantages of a larger constellation size, with latencies becoming less variable as the sizes increase due to the reasons mentioned in the prior experiments.   
\section{Conclusions and Future Work}
\label{sect:future-work}

In terms of the future of the simulation model itself, future work would entail optimising the routing of the model. \cite{handleyUsingGroundRelays2019} uses a similar Dijkstra’s Algorithm routing computation and presents many methodologies in reducing the time complexity of the algorithm. SpaceX plans to have a constellation of 4,408 satellites in its initial phase [18] which will be progressively demanding for the routing computation as the phases advance, making the complexity of the routing computation crucial for simulating later deployments. The simulation model uses the simplistic Unit Disk Radio model described in Section \ref{sec:simulationModel}, which does not model complex satellite characteristics such as modulation and bit error rate. Future work could also involve adapting the INET APSK radio model for more realistic behaviours which was not the focus of the prior simulation experiments. 

Using the developed simulation model, many research areas can now be addressed. The next major step will be to extend and use the simulation model to evaluate the performance of existing data transport protocols. This includes an evaluation of protocols such as TCP \cite{fallSimulationbasedComparisonsTahoe1996}, NDP \cite{handleyReArchitectingDatacenterNetworks2017} and SCDP \cite{alasmarSCDPSystematicRateless2019}. Due to the high fluctuations in latencies, delay-based congestion control may be insufficient, which will also be evaluated using the simulation model.

\label{sect:bib}
\bibliographystyle{plain}
\bibliography{summitReferences}

\begin{thebibliography}{10}

\bibitem{CorningSMF28Ultra}
{{Corning}}\textregistered{} {{SMF}}-28\textregistered{} {{Ultra Optical
  Fiber}}.
\newblock page~2.

\bibitem{FCCForm2018}
{{FCC Form}} - 2018 {{Attachment Technical Informatio
  SAT}}-{{MOD}}-20181108-00083.
\newblock https://fcc.report/IBFS/SAT-MOD-20181108-00083/1569860.

\bibitem{FCCFormAttachment}
{{FCC Form}} - {{Attachment Technical Attach SAT}}-{{MOD}}-20200417-00037.
\newblock https://fcc.report/IBFS/SAT-MOD-20200417-00037/2274316.

\bibitem{GlobalPingStatistics}
Global {{Ping Statistics}}.
\newblock https://wondernetwork.com/pings.

\bibitem{INETFrameworkINET}
{{INET Framework}} - {{INET Framework}}.
\newblock https://inet.omnetpp.org/.

\bibitem{LEOSatellitesTelesat}
{{LEO Satellites}} | {{Telesat}}.
\newblock https://www.telesat.com/leo-satellites/.

\bibitem{OMNeT}
{{OMNeT}}++.
\newblock https://omnetpp.org/.

\bibitem{OneWebOneWorld}
{{OneWeb}} | {{One World}}.
\newblock https://www.oneweb.world/.

\bibitem{SpaceXNongeostationarySatellite}
{{SpaceX}} non-geostationary satellite system {{Attachment A}}: Technical
  information to supplement {{Schedule S}}.

\bibitem{StarlinkSatelliteTrackera}
Starlink satellite tracker.
\newblock https://satellitemap.space.

\bibitem{alasmarSCDPSystematicRateless2019}
Mohammed Alasmar, George Parisis, and Jon Crowcroft.
\newblock {{SCDP}}: {{Systematic Rateless Coding}} for {{Efficient Data
  Transport}} in {{Data Centres}}.
\newblock {\em arXiv:1909.08928 [cs]}, September 2019.

\bibitem{bhattacherjeeGearing21stCentury2018}
Debopam Bhattacherjee, Waqar Aqeel, Ilker~Nadi Bozkurt, Anthony Aguirre,
  Balakrishnan Chandrasekaran, P.~Brighten Godfrey, Gregory Laughlin, Bruce
  Maggs, and Ankit Singla.
\newblock Gearing up for the 21st century space race.
\newblock In {\em Proceedings of the 17th {{ACM Workshop}} on {{Hot Topics}} in
  {{Networks}}}, pages 113--119, {Redmond WA USA}, November 2018. {ACM}.

\bibitem{boleySatelliteMegaconstellationsCreate2021}
Aaron~C. Boley and Michael Byers.
\newblock Satellite mega-constellations create risks in {{Low Earth Orbit}},
  the atmosphere and on {{Earth}}.
\newblock {\em Scientific Reports}, 11(1):10642, May 2021.

\bibitem{delaneyProblemProjections}
Ian Delaney.
\newblock The problem with projections.
\newblock https://360.here.com/2015/05/25/problem-projections/.

\bibitem{fallSimulationbasedComparisonsTahoe1996}
Kevin Fall and Sally Floyd.
\newblock Simulation-based comparisons of tahoe, reno and {{SACK TCP}}.
\newblock {\em SIGCOMM Comput. Commun. Rev.}, 26(3):5--21, July 1996.

\bibitem{handleyDelayNotOption2018}
Mark Handley.
\newblock Delay is {{Not}} an {{Option}}: {{Low Latency Routing}} in {{Space}}.
\newblock In {\em Proceedings of the 17th {{ACM Workshop}} on {{Hot Topics}} in
  {{Networks}}}, pages 85--91, {Redmond WA USA}, November 2018. {ACM}.

\bibitem{handleyUsingGroundRelays2019}
Mark Handley.
\newblock Using ground relays for low-latency wide-area routing in
  megaconstellations.
\newblock In {\em Proceedings of the 18th {{ACM}} Workshop on Hot Topics in
  Networks}, {{HotNets}} '19, pages 125--132, {New York, NY, USA}, 2019.
  {Association for Computing Machinery}.

\bibitem{handleyReArchitectingDatacenterNetworks2017}
Mark Handley, Costin Raiciu, Alexandru Agache, Andrei Voinescu, Andrew~W.
  Moore, Gianni Antichi, and Marcin W{\'o}jcik.
\newblock Re-architecting datacenter networks and stacks for low latency and
  high performance.
\newblock In {\em Proceedings of the {{Conference}} of the {{ACM Special
  Interest Group}} on {{Data Communication}}}, {{SIGCOMM}} '17, pages 29--42,
  {New York, NY, USA}, August 2017. {Association for Computing Machinery}.

\bibitem{hootsSpaceTrackReportNo1980}
Felix~R. Hoots and Ronald~L. Roehrich.
\newblock {{SpaceTrack Report No}}. 3 : {{Models}} for {{Propagation}} of
  {{NORAD Element Sets}}.
\newblock Technical report, {Defense Technical Information Center}, {Fort
  Belvoir, VA}, December 1980.

\bibitem{klenzeNetworkingHeavenEarth2018}
Tobias Klenze, Giacomo Giuliari, Christos Pappas, Adrian Perrig, and David
  Basin.
\newblock Networking in {{Heaven}} as on {{Earth}}.
\newblock In {\em Proceedings of the 17th {{ACM Workshop}} on {{Hot Topics}} in
  {{Networks}}}, pages 22--28, {Redmond WA USA}, November 2018. {ACM}.

\bibitem{niehoeferCNIOpenSource2013}
Brian Niehoefer, Sebastian {\v S}ubik, and Christian Wietfeld.
\newblock The {{CNI}} open source satellite simulator based on {{OMNeT}}++.
\newblock January 2013.

\bibitem{spacexStarlink}
SpaceX.
\newblock Starlink.
\newblock https://www.starlink.com.

\end{thebibliography}


\end{document}